\documentclass[aps,prb,showpacs,reprint,superscriptaddress]{revtex4-1}

\usepackage{graphicx}
\usepackage{color}
\usepackage{amsmath}
\usepackage{bbm}
\usepackage{amssymb}

\usepackage{dsfont}

\usepackage[utf8]{inputenc}	
\usepackage[T1]{fontenc}

\input epsf

\begin{document}

\title{Superconductivity in the three-band model of cuprates: Variational wave function study and relation to the single-band case}
\author{M. Zegrodnik}
\email{michal.zegrodnik@agh.edu.pl}
\affiliation{Academic Centre for Materials and Nanotechnology, AGH University of Science and Technology, Al. Mickiewicza 30, 30-059 Krakow,
Poland}
\author{A. Biborski}
\email{andrzej.biborski@agh.edu.pl}
\affiliation{Academic Centre for Materials and Nanotechnology, AGH University of Science and Technology, Al. Mickiewicza 30, 30-059 Krakow,
Poland}
\author{M. Fidrysiak}
\email{maciej.fidrysiak@uj.edu.pl}
\affiliation{Marian Smoluchowski Institute of Physics, 
Jagiellonian University, ul. \L ojasiewicza 11,
30-348 Krakow, Poland}
\author{J. Spa\l ek}
\email{jozef.spalek@uj.edu.pl}
\affiliation{Marian Smoluchowski Institute of Physics, 
Jagiellonian University, ul. \L ojasiewicza 11,
30-348 Krakow, Poland}

\date{04.12.2018}

\begin{abstract}
The $d$-$wave$ superconductivity is analyzed within the three-band $d$-$p$ model with the use of the diagrammatic expansion of the Guztwiller wave function method (DE-GWF). The determined stability regime of the superconducting state appears in the range of hole doping $\delta\lesssim 0.35$, with the optimal doping close to $\delta\approx 0.19$. The pairing amplitudes between the $d$-orbitals due to copper and $p_x/p_y$ orbitals due to oxygen are analyzed together with the hybrid $d$-$p$ pairing. The $d$-$d$ pairing between the nearest neighboring atomic sites leads to the dominant contribution to the SC phase. Moreover, it is shown that the decrease of both the Coulomb repulsion on the copper atomic sites ($U_d$) and the charge transfer energy between the oxygen and copper atomic sites ($\epsilon_{dp}$) increases the pairing strength as it moves the system from the strong to the intermediate-correlation regime, where the pairing is maximized. Such a result is consistent with our analysis of the ratio of changes in the hole content at the $d$ and $p$ orbitals due to doping, which, according to experimental study, increases with the increasing maximal critical temperature [cf. Nat. Commun. \textbf{7}, 11413 (2016)]. Furthermore, the results for the three-band model are compared to those for the effective single-band picture and similarities between the two approaches are discussed. For the sake of completeness, the normal-state characteristics determined from the DE-GWF approach are compared with those resulting from the Variational Quantum Monte Carlo method with inter-site correlations included through the appropriate Jastrow factors.
\end{abstract}

\maketitle
\section{Introduction}
A complete theoretical description of unconventional superconductivity (SC) in the copper-based materials has long been the subject of debate and still remains an open issue. The main question concerns the microscopic mechanism which can lead to the high-temperature superconductivity, as well as the determination of a proper minimal model which would capture its principal properties. Since the cuprates belong to the group of strongly correlated electron systems, the application of standard DFT ab-initio calculations seems questionable. On the other hand, methods dedicated specifically to the description of strong electron correlations are involved and their application to significantly simplified models only appears to be feasible so far. 

It is believed that the copper-oxide planes which are common to the whole cuprate family are instrumental for the formation of the SC phase when the antiferromagnetic charge transfer insulating parent compound is doped with electrons or holes\cite{Anderson1988,cuprates_rev_2006,ZRS1988}. Therefore, a significant effort has been devoted to determine which of the copper and oxygen orbitals should be taken into account in the appropriate minimal model. The simplest and most commonly used approach incorporates the copper and oxygen degrees of freedom into a single-band picture with the Zhang-Rice singlets (ZRS)\cite{ZRS1988} playing the role of quasiparticles. In this respect, both Hubbard and $t$-$J$ models have been intensively investigated\cite{tJmodel_rev_2008,Dagotto1994}. Within such an approach the Coulomb repulsion is regarded as the largest parameter in the system what leads to the SC phase due to strong electronic correlations without any explicitly attractive interaction. However, alternative approaches within the weak coupling scenario, such as the spin-fluctuation induced pairing, also have been discussed\cite{spin_fluct_rev}, partially in the form adopted to the strong-correlation regime\cite{Plakida2016}. The strong correlation induced SC phase within the $t$-$J$ model occurs already at the renormalized mean-field theory level\cite{Gros2007}, whereas for the case of the Hubbard model one has to include the correlation effects beyond the RMFT for the pairing to occur \cite{VMC_Hubbard_2007, DMFT_SC_Hub_2006, Hub_SC_2005}. The single-band approach which combines the features of both $t$-$J$ and Hubbard models, is the so-called $t$-$J$-$U$ model\cite{Zegrodnik_1,Zegrodnik_2,Zegrodnik_3,Zegrodnik_4,Zegrodnik_5,Fidrysiak2018}. For the latter we have obtained very good agreement between theoretical results and the principal experimental observations concerning the pure $d$-wave SC state\cite{Zegrodnik_1}.

In spite of definite successes of the single-band picture, particular factors influencing the correlation-induced SC state in the Cu-O planes still have not been resolved within that approach.  Namely, the doped holes preferentially reside on the oxygen orbitals and a proper partition of the carriers among the Cu and O atoms seems to be essential in maximizing the $T_C$\cite{Zheng1995,Rybicki2016}. In connection to that it has been also argued that the value of the maximal critical temperature is significantly influenced by the value of the charge transfer gap\cite{Weber_epl2012,Ruan_Sci_Biull2016}, which is determined by the energy distance between the copper and oxygen atomic-levels. Under this circumstances an explicit inclusion of the oxygen degrees of freedom should be considered as an important ingredient to be included in any minimal model of the hole-doping-induced superconductivity. The simplest model which takes this into account is the three-band model consisting of the $3d_{x^2-y^2}$ orbital due to copper hybridized with $2p_x/2p_y$ orbitals due to oxygen. So far, the application of the dynamical mean field theory\cite{DMFT_3band_2008,3band_PRL_Mott}, variational wave function approach\cite{3band_VMC_Mott}, as well as of determinant quantum Monte Carlo\cite{QMC_3band_2016} methods, has lead to reproduction of the charge transfer insulating phase for the half-filled situation, which corresponds to five valence electrons per CuO$_2^{--}$ complex. Moreover, the appearance of the magnetically ordered (AF/SDW) states close to the half-filling and superconductivity have been studied with the use of variational wave functions\cite{3band_Variational_SDW_SC,3band_variational_AF,3band_variational_SDW_SC_2,Yanagisawa2008}. The dome-like behavior of the SC amplitude as a function of doping as well as the anticorrelation between the charge-transfer energy value and the maximal $T_C$, has been reported with the use of the cluster DMFT calculations\cite{Weber_epl2012}.

Here, we apply the approach based on the Gutzwiller- and Jastrow-type wave functions to study both the selected normal-state characteristics and the paired state within the three-band ($d$-$p$) model. The methods in use: (i) are the diagrammatic expansion of the Guzwiller wave function (DE-GWF) and (ii) the variational quantum Monte Carlo (VMC) with Jastrow correlations. We analyze the pairing amplitudes among the $d$- and $p_x/p_y$-orbitals, as well as the hybrid $d$-$p$ pairing, to determine which orbitals constitute the dominant contribution to the superconducting state. We also show that the Gutzwiller-type variational wave function captures the dome-like behavior of the dominant SC amplitude as a function of hole doping. Furthermore, the influence of the charge transfer energy and the Coulomb repulsion on both the pairing strength and the relative occupancy on the $d$ and $p$ orbitals is discussed in the context of experimental observations for the cuprates\cite{Rybicki2016,Ruan_Sci_Biull2016}. Throughout our analysis we focus on the comparison between the single- and three-band pictures and discuss to what extent the former is efficient in the description of the SC phase by relating directly the corresponding macro properties.

In the following Section we present the details of the theoretical model and the applied calculation methods. In Section III we first analyze the normal state characteristics with the use of both VMC and DE-GWF approaches. Next, we move to the detailed analysis of the paired state within the DE-GWF method and compare our results with the single-band case. The conclusions are deferred to Section IV.



\section{Model and method}
\begin{figure}
 \centering
 \includegraphics[width=0.30\textwidth]{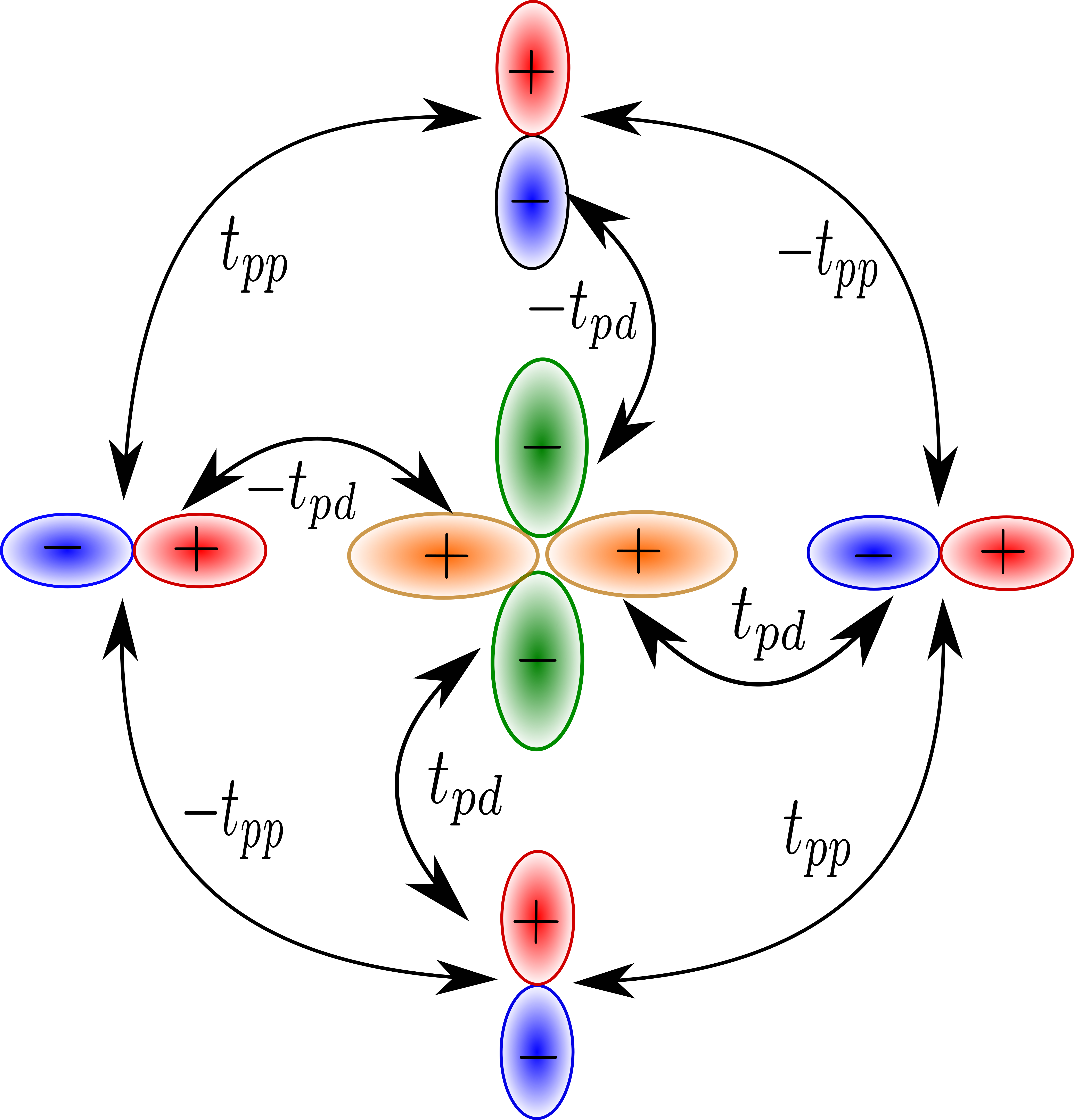}
 \caption{The hopping parameters between the three types of orbitals in the model and the corresponding sign convention for the antibonding orbital structure. The $d_{x^2-y^2}$  orbital is centered at the copper site and the $p_x/p_y$ orbitals are centered at the oxygen sites.}
 \label{fig:Cu_O_hoppings}
\end{figure}

We start from the three-band $d$-$p$ model
\begin{equation}
\begin{split}
 \hat{H}&=\sum_{\langle il,jl'\rangle}t^{ll'}_{il}\hat{c}^{\dagger}_{il\sigma}\hat{c}_{jl'\sigma}+\sum_{il}(\epsilon_{l}-\mu)\hat{n}_{il}+\sum_{il}U_{l}\hat{n}_{il\uparrow}\hat{n}_{il\downarrow}\\ 
 \end{split}
 \label{eq:Hamiltonian_start}
\end{equation}
where $\hat{c}^{\dagger}_{il\sigma}$ ($\hat{c}_{il\sigma}$) creates (anihilates) electron with spin $\sigma$ at the $i$-th atomic site corresponding to orbital denoted by $l\in\{{d,p_x,p_y}\}$ and $\langle il,jl'\rangle$ means that the summation is carried out only for the interorbital nearest neighbor hoppings (cf. Fig. \ref{fig:Cu_O_hoppings}). Note that the $p$ orbitals are located at the oxygen atomic sites which reside in between every two nearest neighbor copper sites containing the $d$ orbital states (cf. Fig. \ref{fig:Cu_O_hoppings}). The phase convention has been taken in the electron representation and is provided in Fig. \ref{fig:Cu_O_hoppings}. The second term of the Hamiltonian corresponds to the $d$ and $p_x/p_y$ atomic levels ($\epsilon_{p_x}=\epsilon_{p_y}\equiv\epsilon_{p}$, $\epsilon_d-\epsilon_p\equiv\epsilon_{dp}$), together with the chemical potential contribution. The interaction parameters $U_d$ and $U_{p_x}=U_{p_y}\equiv U_p$ correspond to the intrasite Coulomb repulsion between two electrons with opposite spins located on the $d$ and $p_x/p_y$ orbitals, respectively. 

Hamiltonian (\ref{eq:Hamiltonian_start}) expresses an effective description of the Cu-O planes of the copper based compounds. The values of the hopping and interaction parameters have been evaluated in earlier analysis within the DFT approach\cite{3b_DFT_1989,3b_DFT_1990}, as well as cluster calculations compared with XPS or Auger measurements\cite{Sawatzky_Auger_1988,Fujimori_XPS_1989,Sawatzky_cluster_1990}. More recent analysis with the use of $ab$ $initio$ GW and DFT combination has lead to similar values of model parameters obtained within a single scheme\cite{Hirayama_GWDFT_2018} which does not suffer from the so-called double counting interaction problem.  

\begin{figure}
 \centering
 \includegraphics[width=0.5\textwidth]{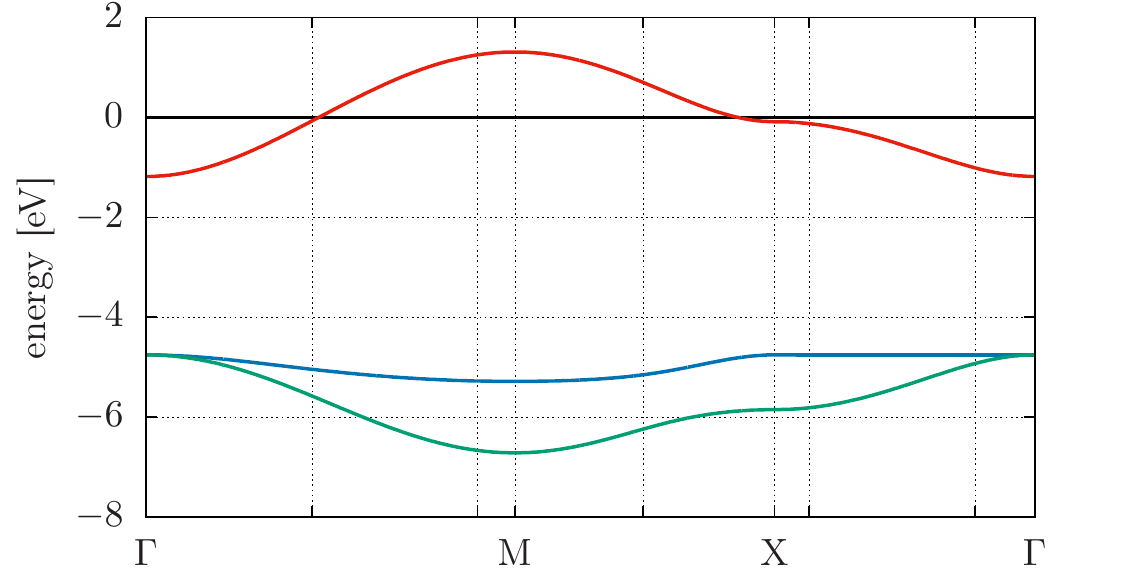}
 \caption{The electronic structure of the single-particle part of Hamiltonian (\ref{eq:Hamiltonian_start}) with parameters: $t_{dp}=1.13\;$eV, $t_{pp}=0.49\;$eV, and $\epsilon_{dp}=3.57\;$eV (cf. Fig. \ref{fig:Cu_O_hoppings}). Fermi energy has been taken as the reference value ($E=0$) on the vertical axis, and corresponds to the case of five electrons per CuO$_2$ complex, which is referred as the half-filled situation.}
 \label{fig:disrel_H0}
\end{figure}

The electronic structure corresponding to the single-particle part of Hamiltonian (\ref{eq:Hamiltonian_start}), with typical values of the bare hopping parameters $t_{dp}=1.13\;$eV, $t_{pp}=0.49\;$eV, and the charge-transfer energy $\epsilon_{dp}=3.57\;$eV, is shown in Fig. \ref{fig:disrel_H0} and consists of hybridized $dp$ antibonding band (red solid line), which crosses the Fermi surface and two fully filled low energy bands (blue and green solid lines). The typical values of the interaction parameter $U_d$ ($U_p$) range between $8-10.5\;$eV ($4-6\;$eV), depending on the particular approach\cite{3b_DFT_1989,3b_DFT_1990,Hirayama_GWDFT_2018}. As the value of $U_d$ is significant, the system should be analyzed with the use of a method dedicated to capture the many-body effects resulting from strong electronic correlations. In our analysis we use two methods, which are based on the variational wave functions. Namely, the DE-GWF method which allows us to determine the full Gutzwiller wave function solution for an infinite system, and the VMC approach applied for system of limited size with both the on-site Gutzwiller and intersite Jastrow factors included. To emphasize the effect of strong electronic correlations, the Hartree-Fock results are also provided for comparison.

\subsection{Three-band d-p model within the diagrammatic expansion of the Gutzwiller wave function approach}

The description of the DE-GWF method as applied to the analysis of the SC phase within the single-band $t$-$J$, Hubbard, and $t$-$J$-$U$ models is provided in Refs.\onlinecite{Zegrodnik_1,Kaczmarczyk_2013,Kaczmarczyk2014}. The method has been recently used to study SC in the Anderson lattice model\cite{MWysokinski_Anderson} (with reference to the heavy-fermion systems) as well as ferromagnetism and Fermi surface deformations in the two-band Hubbard model\cite{2band_Hub_Bunemann}. Here, we show some details of the calculation scheme as applied to the three-band $d$-$p$ model of superconductivity in the cuprates.

The Gutzwiller-type projected many particle wave function is taken in the form
\begin{equation}
 |\Psi_G\rangle\equiv\hat{P}|\Psi_0\rangle=\prod_{il}\hat{P}_{il}|\Psi_0\rangle \;,
 \label{eq:GWF}
\end{equation}
where $|\Psi_0\rangle$ represents the wave function of uncorrelated SC state. The main difference between the present application and that of the single-band case is that here the situation is orbital-dependent, i.e., 
\begin{equation}
 \hat{P}_{il}\equiv \sum_{\Gamma}\lambda_{\Gamma|il}|\Gamma\rangle_{il\;il}\langle\Gamma|\;,
 \label{eq:P_Gamma}
\end{equation}
with $\lambda_{\Gamma|il}$ being the set of variational parameters determining relative weights corresponding to $|\Gamma\rangle_{il}$, which represent states from the local basis on the atomic sites with the three types of orbitals ($l\in\{{d,p_x,p_y}\}$)
\begin{equation}
|\Gamma\rangle_{il}\in \{|\varnothing\rangle_{il}, |\uparrow\rangle_{il}, |\downarrow\rangle_{il},
|\uparrow\downarrow\rangle_{il}\}\;.
\label{eq:local_states}
\end{equation}
The consecutive states represent the empty, singly, and doubly occupied local configurations, respectively. As can be seen, the variational parameters, which tune the local electronic configurations in the resulting wave function, are orbital-dependent. 

To simplify significantly the calculations and improve the convergence, one can impose the condition\cite{Bunemann2012,Gebhard1990} 
\begin{equation}
 \hat{P}_{il}^2\equiv 1+x_{il}\hat{d}^{\textrm{HF}}_{il}\;,
 \label{eq:condition}
\end{equation}
where $x_{il}$ is yet another variational parameter and $\hat{d}^{\textrm{HF}}_{il}=\hat{n}_{il\uparrow}^{\textrm{HF}}\hat{n}_{il\downarrow}^{\textrm{HF}}$, $\hat{n}_{il\sigma}^{\textrm{HF}}\equiv\hat{n}_{il\sigma}-n_{l0}$, 
with $n_{l0}\equiv\langle\Psi_0|\hat{n}_{il\sigma}|\Psi_0\rangle$. By using Eqs. (\ref{eq:P_Gamma}) and (\ref{eq:condition})
one can express the parameters $\lambda_{\Gamma|il}$ with the use of $x_{il}$. Since we are considering a spatially homogeneous state the $i$ indices in the variational parameters $\lambda_{\Gamma|il}$ and $x_{il}$ can be dropped. Moreover, the oxygen orbitals $p_x$ and $p_y$ are equivalent which allows us to take $\lambda_{\Gamma|p_x}=\lambda_{\Gamma|p_y}\equiv\lambda_{\Gamma|p}$ and $x_{p_x}=x_{p_y}\equiv x_p$. However, the Coulomb repulsion $U_d$ is diferent from $U_p$ and the occupancy of the $d$ and $p$ orbitals also, which means that $\lambda_{\Gamma|d}\neq\lambda_{\Gamma|p}$ and $x_d\neq x_p$. In order to simplify further the calculations one can take $x_p\equiv 0$, for which $\hat{P}_{ip_x}=\hat{P}_{ip_y}\equiv\mathds{1}$ and then only the copper atomic sites are affected by the correlation operator $\hat{P}$. Such approximation is justified by the fact that the Coulomb repulsion among the $p$ orbitals is significantly weaker then that corresponding to $d$ orbitals ($U_d>U_p$). In Appendix A we show that for the parameter regime appropriate for the cuprates such approximation does not change significantly the obtained results.

 In the next step one has to express the expectation values of all the terms appearing in Hamiltonian (\ref{eq:Hamiltonian_start}) in the correlated state $|\Psi_G\rangle$. For example, for the hopping term the corresponding expectation value takes the form
 \begin{equation}
 \begin{split}
  \langle\Psi_G|&\hat{c}^{\dagger}_{il\sigma}\hat{c}_{jl'\sigma}|\Psi_G\rangle=\\
  &\sum_{k=0}^{\infty}\frac{1}{k!}\sideset{}{'}\sum_{m_1f_1...m_kf_k}x^{k_d}_dx^{k_p}_p
  \langle\tilde{c}^{\dagger}_{il\sigma}\tilde{c}_{jl\sigma}\hat{d}^{\textrm{HF}}_{m_1f_1}...\hat{d}^{\textrm{HF}}_{m_kf_k} \rangle_0\;,
  \end{split}
  \label{eq:diag_sum}
 \end{equation}
where $\hat{d}^{\textrm{HF}}_{\varnothing}\equiv 0$, $\tilde{c}^{(\dagger)}_{il\sigma}\equiv \hat{P}_{il}\hat{c}^{(\dagger)}_{il\sigma}\hat{P}_{il}$, 
and the index $m$ corresponds to lattice sites, whereas $f$ enumerates the orbitals. The primmed summation on the right hand side is restricted to $(l_h,m_h)\neq (l_{h'},m_{h'})$, $(l_h,m_h)\neq (i,l)$, $(l_h,m_h)\neq (j,l')$ for all $h$, $h'$.
The powers $k_d$ ($k_p$) express how many times the indices $f_h$ on the right hand of equation (\ref{eq:diag_sum})
have the value corresponding to $d$ ($p$) orbital. For given $k$, they fulfill the relation $k_d+k_p=k$.  The maximal $k$, for which the terms in Eq. (\ref{eq:diag_sum}) are taken into account represents the order of calculations. Similar expressions can be derived for the case of the Coulomb repulsion terms\cite{Zegrodnik_1}. It has been shown\cite{Bunemann2012, Kaczmarczyk2014,Abram2017} that the first 4-6 terms of the expansion lead to sufficient accuracy of the method. In the subsequent analysis the calculations have been carried out in the third order of the expansion. 

Note that the expectation values on the right-hand-side of Eq. (\ref{eq:diag_sum}) are taken in non-correlated state, $\langle...\rangle_0\equiv\langle\Psi_0|...|\Psi_0\rangle$, what allows us to carry out the Wicks decomposition. As a result, one obtains the system energy in the correlated state expressed in terms of the variational parameters $x_d$, $x_p$ and the non-correlated hoppings and pairing, $P_{ijll''\sigma}\equiv\langle\hat{c}^{\dagger}_{il\sigma}\hat{c}_{j\sigma} \rangle_0$, $S_{ijll'}\equiv\langle\hat{c}^{\dagger}_{il\uparrow}\hat{c}^{\dagger}_{jl'\downarrow} \rangle_0$, respectively.

To determine explicitly the values of $P_{ijll''\sigma}$ and $S_{ijll''\sigma}$, as well as the variational parameters, the 
grand-canonial potential $\mathcal{F}=\langle\hat{H}\rangle_G-\mu_G\langle\hat{n}\rangle_G$ ($\langle\hat{o}\rangle_G\equiv \langle\Psi_G|\hat{o}|\Psi_G\rangle_G/\langle\Psi_G|\Psi_G\rangle_G$ and $\mu_G$ is the chemical potential) is minimized. The minimization condition can be cast into the form of the effective Hamiltonian\cite{Kaczmarczyk_2013,Kaczmarczyk2014}
\begin{equation}
\begin{split}
 \hat{\mathcal{H}}_{\textrm{eff}}&=\sideset{}{'}\sum_{ijll'\sigma}t^{\textrm{eff}}_{ijll'}\hat{c}^{\dagger}_{il\sigma}\hat{c}_{jl'\sigma}+\sum_{il\sigma}\epsilon^{\textrm{eff}}_{il}\hat{n}_{il\sigma}\\
 +&\sum_{ijll'}\big(\Delta^{\textrm{eff}}_{ijll'}\hat{c}^{\dagger}_{il\uparrow}\hat{c}^{\dagger}_{jl'\downarrow}+H.c.\big),
 \end{split}
 \label{eq:H_effective}
\end{equation}
where the primmed summation means $i\neq j$ and the effective hopping, effective superconducting gap, and effective atomic level parameters are defined through the relations 
\begin{equation}
 t^{\textrm{eff}}_{ijll'}\equiv \frac{\partial\mathcal{F}}{\partial P_{ijll'\sigma}},
 \quad \Delta^{\textrm{eff}}_{ijll'}\equiv \frac{\partial\mathcal{F}}{\partial S_{ijll'}},\quad \epsilon^{\textrm{eff}}_{il}\equiv \frac{\partial\mathcal{F}}{\partial n^0_{il\sigma}}.
 \label{eq:effective_param}
\end{equation}
As one can see, the effective Hamiltonian contains both ($l=l'$) intra- and ($l\neq l'$) inter- orbital pairing amplitudes. Nevertheless, all the amplitudes possess the $d$-$wave$ symmetry and no on-site pairing appears. The above real-space representation can be transformed into reciprocal space and diagonalized through the 6x6 generalized Bogoliubov-de Gennes transformation, on the basis of which the self consistent equations for the pairing and hopping expectation values can be derived. Within such a scheme, the minimization over the variational parameters $x_d$ and $x_p$ has to be incorporated into the procedure of solving the self-cosistent equations.

\begin{figure}
 \centering
 \includegraphics[width=0.45\textwidth]{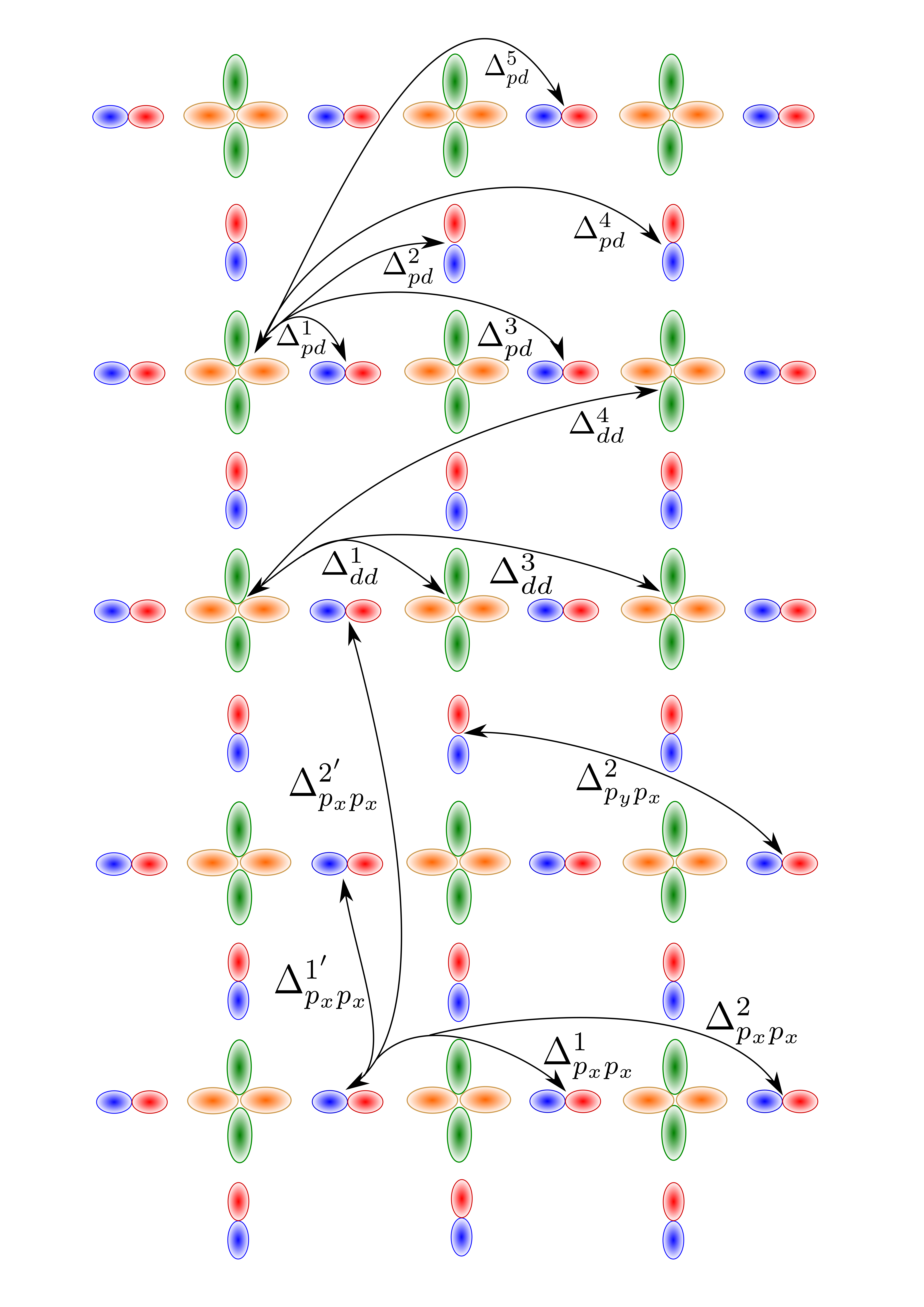}
 \caption{The component pairing amplitudes in the correlated state, $|\Psi_G\rangle$, that are taken into account within our scheme. The superscripts correspond to pairing between consecutive nearest-neighbors of given type ($d$-$d$, $d$-$p$, $p_x$-$p_x$, $p_y$-$p_y$, $p_y$-$p_x$). The effective gap parameters of corresponding types are also analyzed here, but are not marked for the sake of clarity.}
 \label{fig:Cu_O_pairing}
\end{figure}

After all the lines, together with the variational parameters, are determined, one can calculate next the values of superconducting pairing amplitudes between particular atomic sites in the correlated state, $|\Psi_G\rangle$. In the subsequent Section we are going to analyze both the SC pairing amplitudes and the effective gap parameters for the case of intra-orbital ($d$-$d$, $p_x$-$p_x$, and $p_y$-$p_y$) and inter-orbital ($d$-$p_x$, $d$-$p_y$, and $p_x$-$p_y$) pairing between various nearest-neighbors in the Cu-O lattice. The notation is
\begin{equation}
    \Delta^f_{ll'} \equiv \langle\hat{c}^{\dagger}_{il\uparrow}\hat{c}^{\dagger}_{jl'\downarrow}\rangle_G,\quad
     \Delta^{f}_{\textrm{eff},ll'}\equiv \partial\mathcal{F}/\partial S_{ij,ll'},
     \label{eq:SC_amplitudes}
\end{equation}
where the $f$ superscript defines nearest-neighbors of given type. For example, for the case of $d$-$d$ pairing the SC amplitudes and effective gaps up to the fourth neighbor are included $\Delta^1_{dd}$, $\Delta^3_{dd}$, $\Delta^4_{dd}$, $\Delta^1_{\mathbf{eff},dd}$, $\Delta^3_{\mathbf{eff},dd}$, and $\Delta^4_{\mathbf{eff},dd}$. The second $d$-$d$ neighbor is excluded, since we are assuming the $d$-wave symmetry. All the correlated pairing amplitudes taken into account in the calculations are shown in Fig. \ref{fig:Cu_O_pairing}. In our notation $\Delta^{f}_{p_xp_x}$ ($\Delta^{f}_{p_yp_y}$) corresponds to the $p_x$-$p_x$ ($p_y$-$p_y$) pairing in the $(1,0)$ [$(0,1)$] direction, whereas $\Delta^{f'}_{p_xp_x}$ ($\Delta^{f'}_{p_yp_y}$) to the $p_x$-$p_x$ ($p_y$-$p_y$) pairing in the $(0,1)$ [$(1,0)$] direction. The $p_x$-$p_x$ and $p_y$-$p_y$ pairing in the $(1,0)$ and $(0,1)$ direction can have different values due to the orientation of the orbitals. However, the corresponding relation is fulfilled $\Delta^{f}_{p_xp_x}=\Delta^{f'}_{p_yp_y}$. This remark corresponds also to the effective gap parameters which are not marked in Fig. \ref{fig:Cu_O_pairing} for the sake of clarity.

\subsection{Variational Monte-Carlo scheme: Application to the three-band d-p model}
\begin{figure}
    \centering
    \includegraphics[width=0.8\linewidth]{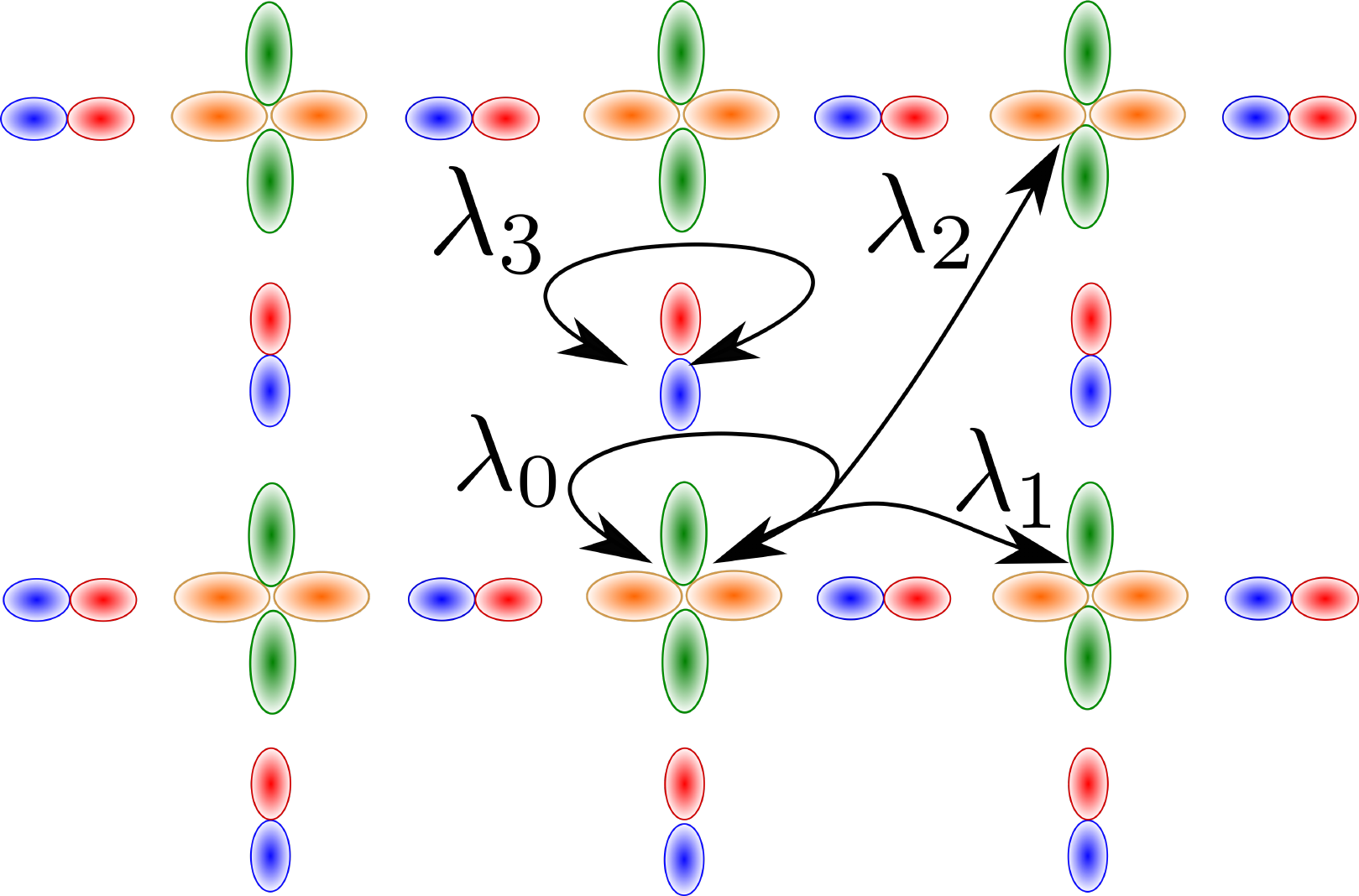}
    \caption{The Jastrow variational parameters considered in the VMC. Site-orbital indicies are re-numbered as follows from the cartoon. }
    \label{fig:jastrows}
\end{figure}

As the supplementary method which validates the results obtained by means of the DE-GWF formalism for the normal (non-SC) state, we exploit the Variational Monte-Carlo approach (VMC). The main advantage of VMC relies on the straightforward inclusion of the inter-electronic correlations in the wave function optimization scheme. In our situation this is performed in the standard manner, i.e., by using the Jastrow correlation operator $\hat{P}_{J}$, defined as 
\begin{equation}
 \hat{P}_{J}\equiv \text{exp}\Big(-\frac{1}{2}\sum_{ij,\mu\nu}\lambda_{ij}^{\mu\nu}\hat{n}_{i}^{\mu}\hat{n}^{\nu}_{j}\Big),
 \label{eq:P_Jastrow}
\end{equation}
(where $\mu,\nu \in \{d,p\}$) acts on the non-correlated state $|\Psi_0\rangle$, and, $\{\lambda_{ij}^{\mu\nu}\}$ are the variational parameters, which are optimized via the VMC scheme.  The details of the VMC procedure may be found in Refs. \onlinecite{Becca,Foulkes}. Here, we have employed the self-developed code ~\cite{qmtURL}, which has been recently applied in a different context ~\cite{Biborski2}.  Formerly ~\cite{Biborski2}, we decided to exploit \emph{variance optimization}~\cite{Becca}, however, ~\emph{energy} optimization is considered to be more robust when \emph{stochastic reconfiguration} ~\cite{Becca} (SR) technique is utilized. Therefore, all the presented results obtained by means of VMC refer to SR based optimization procedure. As VMC operates in real space, the considered systems are \emph{finite} clusters. To minimize the influence of this factor, we have imposed periodic boundary conditions (PBC) in our calculations. We have selected variational parameters to obtain the essential properties of the normal state. Moreover, we found out  that inclusion of particular type of parameters influences the numerical stability of the optimization procedure. This is the case for nearest neighbor $d$-$p$ variational parameter, $\lambda_{ij}^{dp}$. Eventually, after number of testing simulations we decided to limit ourselves to the parameters presented in \ref{fig:jastrows} (note that we provided the re-numbering of Jastrow variational parameters for the sake of brevity). We find this selection as the  compromise between reliability (i.e., trial wave-function flexibility) and numerical stability.

\section{Results}
In this Section we study both the principal normal state characteristics and $d$-$wave$ superconductivity for the case of three-band $d$-$p$ model with either electron or hole doping. In all the figures the zero doping ($\delta=0$) case corresponds to the parent compound for which each CuO$_2$ complex is occupied by five electrons ($n_{tot}=5$). Such a situation is going to be referred as that of the half filling. The $\delta>0$ ($\delta<0$) situation refers to hole (electron) doping with $n_{tot}<5$ ($n_{tot}>5$). If not stated otherwise, we set the hopping parameters and charge transfer energy to $t_{dp}=1.13\;$eV, $t_{pp}=0.49\;$eV, and $\epsilon_{dp}=3.57\;$eV. The interaction parameters $U_d$ and $U_p$ are specified explicitly in each particular case analyzed.

\subsection{Normal state characteristics}

\begin{figure}[h!]
 \centering
 \includegraphics[width=0.50\textwidth]{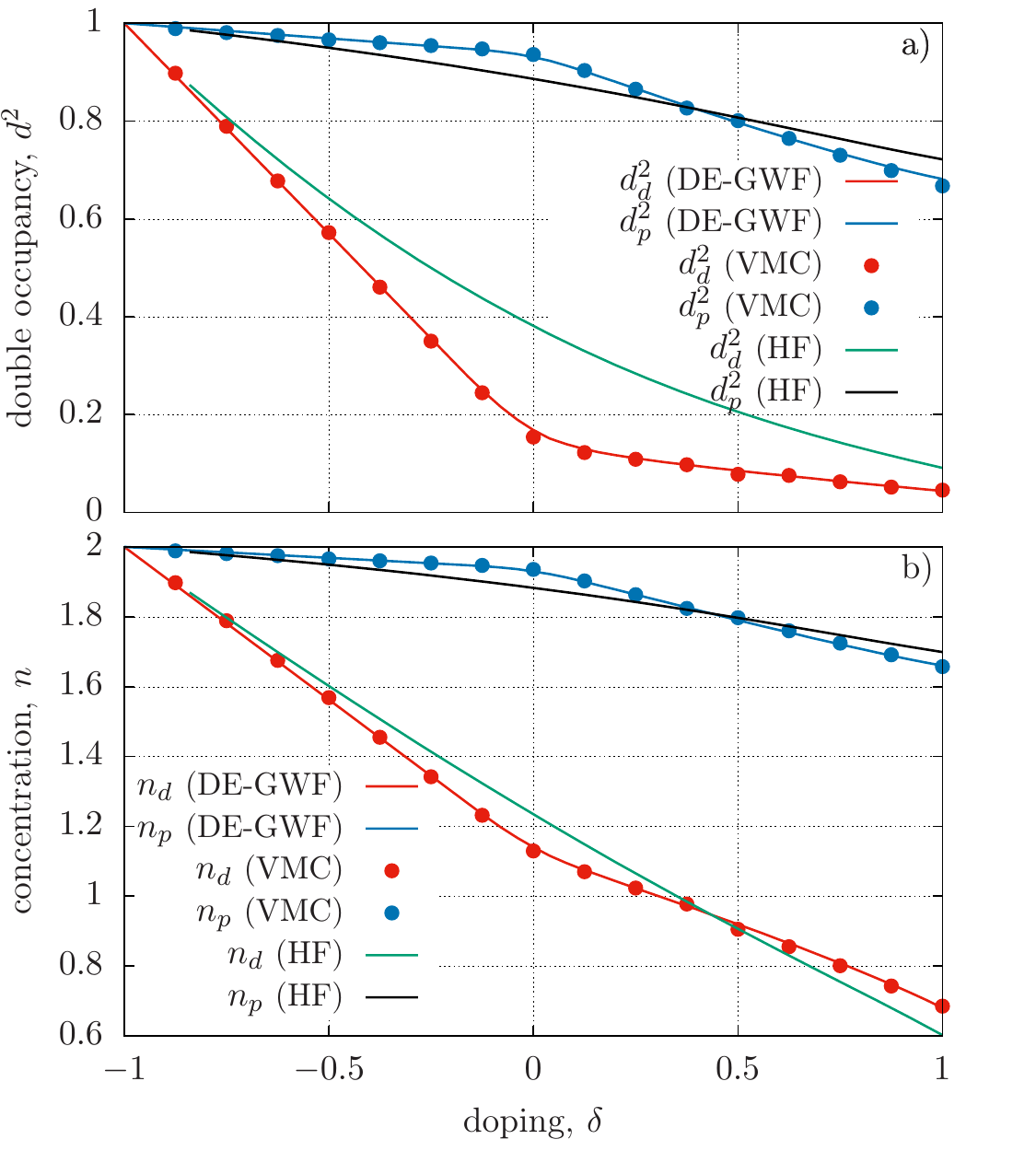}
 \caption{Orbital resolved double occupancies (a) and component electron concentrations (b) as a function of doping for $U_d=7.85\;$eV and $U_p=4.1\;$eV, for three different methods: DE-GWF, VMC, and HF. For $\delta>0$ ($\delta<0$) we have the hole- (electron-) doped case with $n_{tot}<5$ ($n_{tot}>5$).}
 \label{fig:PM_characteristics}
\end{figure}

\begin{figure}
 \centering
 \includegraphics[width=0.5\textwidth]{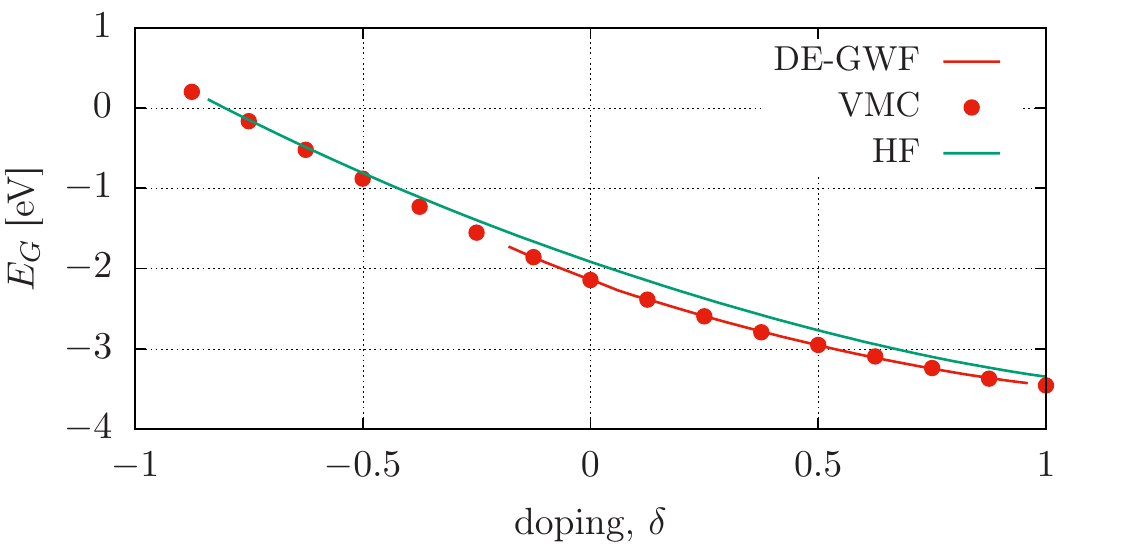}
 \caption{Ground state energy as a function of doping for $U_d=7.85\;$eV and $U_p=4.1\;$eV and for three different methods: DE-GWF, VMC, and HF. For $\delta>0$ ($\delta<0$) we have the hole- (electron-) doped case with $n_{tot}<5$ ($n_{tot}>5$).}
 \label{fig:PM_energy}
\end{figure}

In Figs. \ref{fig:PM_characteristics} and \ref{fig:PM_energy} we show orbital resolved double occupancies ($d_d^2=\langle\hat{n}_{id\uparrow}\hat{n}_{id\downarrow}\rangle_G$, $d_p^2=\langle\hat{n}_{ip_x\uparrow}\hat{n}_{ip_x\downarrow}\rangle_G=\langle\hat{n}_{ip_y\uparrow}\hat{n}_{ip_y\downarrow}\rangle_G$), electron concentrations $n_d$ and $n_p$, as well as the ground state energy, all as a function of doping. For comparison, the calcuations have been carried out by the DE-GWF and VMC methods which take into account correlation effects, as well as the Hartree-Fock (HF) approximation which neglects them. It should be noted that the VMC method includes both intra- and inter-side correlation operators for the case of limited system size of 4x4 CuO$_2$ complexes, whereas within the DE-GWF approach we carry out calculations for an infinite system. However, for the latter approach only on-site Gutzwiller operator for the copper atomic-sites is introduced (cf. Sec. 2). 
As one can see, in spite of the differences both methods provide very similar results with a characteristic kink in both the double occupancies and electron concentration appearing at the half-filling. Within the Hartree-Fock approach such a kink is not reported and a smooth behavior is observed as one passes trough $\delta=0$ point indicating that the kinks result from correlations. Obviously, the HF calculations lead to a visibly higher system energy than the DE-GWF and VMC methods (cf. Fig. \ref{fig:PM_energy}). As one can see, in the hole doped range ($\delta>0$) we have $n_d\lesssim 1$ and due to the high value of $U_d$ the double occupancies at $d$ orbitals are kept relatively small and weakly dependant on the doping. As we increase the number of particles above $n_{tot}=5$ (electron doping, $\delta<0$) the oxygen orbitals are almost completely full, $n_p\approx 2$ with $d_p^2\approx 1$, and the additional doped electrons are forced to occupy the copper orbitals, resulting in visible change of slope at $d^2_d$ and $n_d$. Similar effect, obtained here by the use of variational wave functions, has also been reported within the determinant quantum Monte Carlo approach\cite{QMC_3band_2016} as well as DMFT calculations\cite{DMFT_3band_2008}.

\begin{figure}
 \centering
 \includegraphics[width=0.50\textwidth]{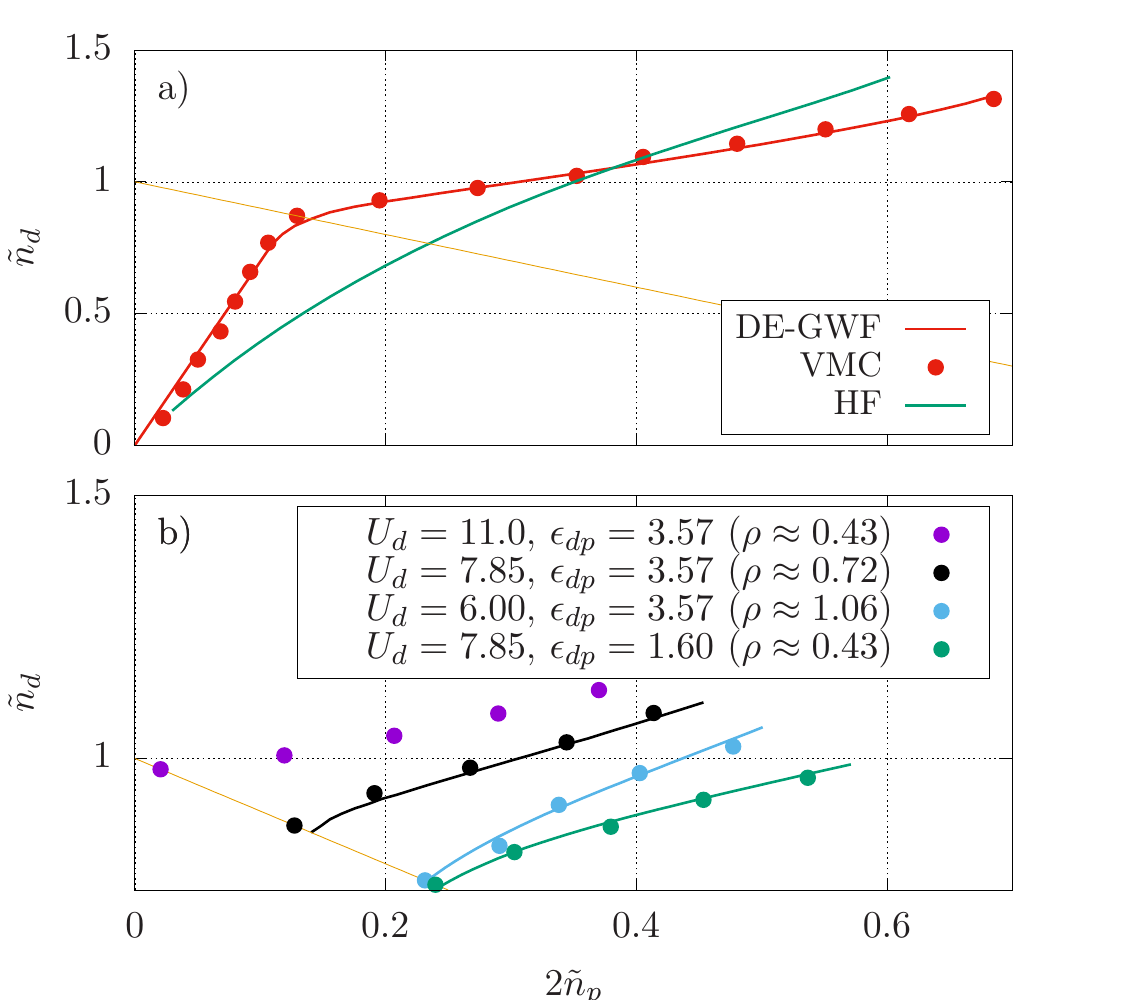}
 \caption{(a) Hole content distribution in the $(\tilde{n}_d,2\tilde{n}_p)$ plane calculated according o the DE-GWF, VMC, and HF methods for $U_d=7.85$ and $U_p=4.1\;$eV. Different points of the plots refer to different values of doping ($\delta$). The solid yellow line corresponds to the parent compound $\delta=0$, the upper-right (lower-left) half of the plane refers to the hole-doped (electron-doped) case. (b) Hole content distribution between the $d$ and $p$ orbitals for different sets of model parameters for the hole-doped situation within the DE-GWF (solid lines) and VMC (circles) methods. For each set the charge distribution ratio, $\rho=\Delta \tilde{n}_d/2\Delta \tilde{n}_p$, has been determined by fitting a linear plot to the VMC data.}
 \label{fig:PM_npnd}
\end{figure}

It has been argued in Ref. \onlinecite{Rybicki2016} that the ratio of respective changes in the hole content on the $d$ and $p$ orbitals ($\rho=\Delta \tilde{n}_d/2\Delta \tilde{n}_p$, where $\tilde{n}_d=2-n_d$, $\tilde{n}_p=2-n_p$), as one increases the number of carriers, is a family property (cf. Fig 2 in Ref. \onlinecite{Rybicki2016}). Moreover, it appears that for the case of hole-doped compounds the smaller $\rho$ is, the smaller the maximal critical temperature of a given group of compounds. In Fig. \ref{fig:PM_npnd} we draw the $\tilde{n}_d$ vs. $\tilde{n}_p$ plot which illustrates the local charge distribution on the $(\tilde{n}_d,2\tilde{n}_p)$ plane. The $\rho$ parameter within the hole- or electron-doping regime can be extracted from the angle between the $\tilde{n}_d(2\tilde{n}_p)$ plot and the horizontal axis. The yellow solid line in the Figure represents the parent compound (one hole per CuO$_2$ complex when $\tilde{n}_d+2\tilde{n}_p=1$), the upper-right half of the $(2\tilde{n}_p, \tilde{n}_d)$ plane corresponds to hole doping, and the lower-left to electron doping. As one can see by comparing the DE-GWF/VMC results with those corresponding to HF approach, the value of $\rho$ in the hole-doped regime is significantly suppressed by the correlation effects taken into account by the variational wave functions. Namely, for the DE-GWF/VMC results we obtain $\rho\approx0.72$, while for HF the result is $\rho\approx 2.0$. The experimental values are in the regime $\rho<1$\cite{Rybicki2016}, reaching even $0.2$ for La-214. We conclude that the low values of $\rho$ for the hole doped cuprates is a signature of strong electron correlations.

In Fig. \ref{fig:PM_npnd} (b) we show the $\tilde{n}_d(2\tilde{n}_p)$ dependence in the hole doped region for different values of model parameters according to DE-GWF (solid line) and VMC (circles). For each set of parameters the approximate value of $\rho$ is determined by fitting the linear plot to our VMC results. As one can see, reduction of $\epsilon_{dp}$ by $2\;$eV, with all the other parameters fixed, does not lead to a significant increase of $\rho$ (black and green data set). Nevertheless, it shifts the plot towards lower $\tilde{n}_d$ and higher $\tilde{n}_p$ values. A similar result but with an additional increase of the $\rho$ parameter is obtained by lowering the $U_d$ value (violet, black, and blue data sets). As shown experimentally both these effects are related to an enhancement of the critical temperature in the cuprates\cite{Rybicki2016}. Obviously, in a realistic situation $U_d$ and $\epsilon_{dp}$ vary between different compounds. According to the recent $ab$ $initio$ calculations\cite{Hirayama_GWDFT_2018} for the two systems with significantly different maximal critical temperatures (HgBa$_2$CuO$_4$ with $T_C\approx 90K$ and La$_2$CuO$_4$ with $T_C\approx 40K$), both the lower $U_d$ and $\epsilon_{dp}$ value correspond to the compound with a higher maximal $T_{C}$. This issue is going to be discussed further in the next subsections, where the paired phase is analyzed in detail. 



\subsection{Superconducting gaps in the three-band model and their single-band correspondant}

\begin{figure}
 \centering
 \includegraphics[width=0.5\textwidth]{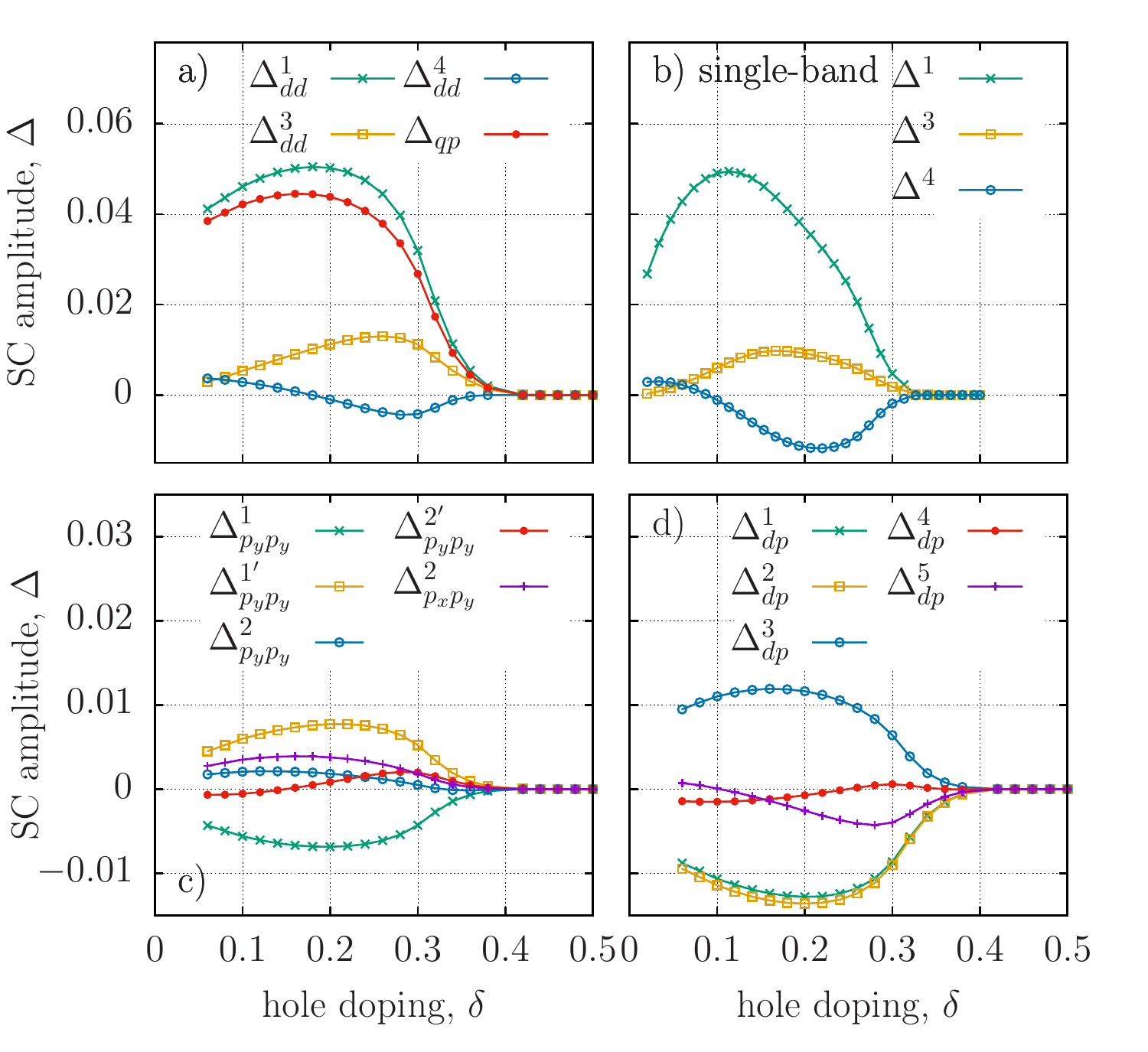}
 \caption{Pairing amplitudes between $d$-$d$ (a), $p$-$p$ (c), and $d$-$p$ (d) atomic sites as a function of doping for $U_d=10.3\;$eV, $U_p=4.1\;$eV (cf. Fig. \ref{fig:Cu_O_pairing} and Eq. (\ref{eq:SC_amplitudes})). Additionally, in (a) we show the quasiparticle gap amplitude ($\Delta_{qp}$) which is defined i the main text. In (b) we show the first, third, and fourth nearest-neighbor pairing amplitudes for the case of single band Hubbard model with $t=-0.35\;$eV, $t'=0.25|t|$, and $U=6\;$eV.}
 \label{fig:delt_tab}
\end{figure}

In this subsection we focus on the analysis of paired state within the three-band model for the case of hole doping ($\delta>0$) within the DE-GWF method. In Fig. \ref{fig:delt_tab} a, c, d we show the doping dependences of the intra- and inter-orbital pairing amplitudes in the correlated state [cf. Fig. \ref{fig:Cu_O_pairing} and Eq. (\ref{eq:SC_amplitudes})]. As one can see, the dominant contribution to the superconducting state results from the pairing between the $d$-orbitals residing on the nearest-neighbor copper atomic sites ($\Delta_{dd}^1$ in Fig. \ref{fig:delt_tab}a). The $\Delta^1_{dd}(\delta)$ function reproduces the dome-like behavior with the maximum value corresponding to the optimal doping, $\delta\approx 0.19$. The maximal values of all the other pairing amplitudes seen in Fig. \ref{fig:delt_tab} a, c, d are approximately $20\%$ of that corresponding to $\Delta_{dd}^1$. It should be noted that even though the nearest-neighbor mixed $d$-$p$ ($\Delta^1_{dp}$) amplitude corresponds to pairing between atomic sites which are twice as close as the ones for the case of $\Delta^1_{dd}$, the latter plays the most important role. It is due to the fact that large $U_{d}$ generates the electron correlations which in turn lead to the paired state. Therefore, the nearest-neighbor sites with the largest Coulomb repulsion constitute the dominant contribution.

For the sake of comparison, in Fig. \ref{fig:delt_tab} b, we show the analogous results for the case of single-band Hubbard model on a square lattice with typical values of model parameters corresponding to the cuprates: $t=-0.35\;$eV, $t'=0.25|t|$, and $U=6\;$eV, which refer to the nearest- and next-nearest neighbor hopping and the on-site Coulomb repulsion, respectivelly. In this case we have only one type of orbital on each atomic site, and we plot the SC amplitudes for the first, third, and fourth nearest-neighbors (second is zero due to the $d$-$wave$ symmetry of the SC state). Note that, the doping dependences of the SC amplitudes are quite similar to the corresponding ones between the $d$ orbitals in the three-band case (cf. Figs. \ref{fig:delt_tab} a and b). These two figures speak for the validity of the single-band picture of the SC gaps.

Additionally, in the three-band case we calculate the SC amplitude $\Delta_{qp}=\langle\hat{\alpha}^{\dagger}_{i\uparrow}\hat{\alpha}^{\dagger}_{j\downarrow}\rangle_G$ (Fig. \ref{fig:delt_tab}a), where $\hat{\alpha}^{\dagger}_{i\sigma}$ are the quasiparticle operators for the hybridized antibonding band which crosses the Fermi surface in the normal state (red solid line in Fig. \ref{fig:disrel_H0}). This is the most representative pairing amplitude in the three-band model since the SC gap is formed around the Fermi surface and the mentioned hybridized band becomes gapped in the SC phase. Since $\Delta^1_{dd}$ has the dominant contribution to the paired state, the behavior of $\Delta_{qp}$ is mostly determined by the former, what can be clearly seen in Fig. \ref{fig:delt_tab}a. 

\begin{figure}
 \centering
 \includegraphics[width=0.5\textwidth]{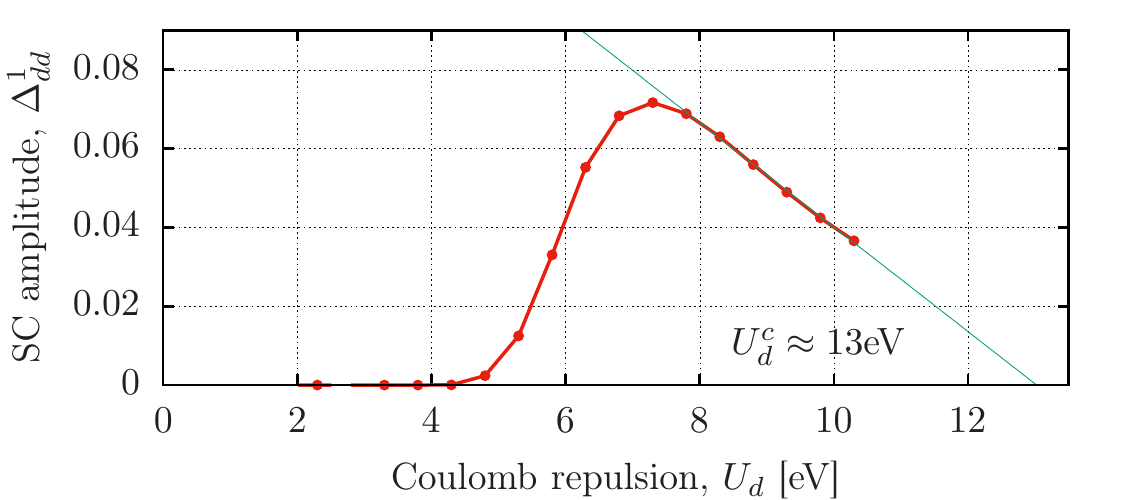}
 \caption{The nearest-neighbor $d$-$d$ pairing amplitude for the case of half-filling, $\delta=0$, as a function of $U_d$ for $U_p=4.1\;$eV. For $U_d\gtrsim 10\;$eV the convergence could not be reached and the critical value of $U_d$ for which $\Delta_{dd}$ is suppressed is evaluated by carrying out a linear plot fitting and leads to the critical value $U_{d}^c\approx 13\;$eV.}
 \label{fig:delt_Udep}
\end{figure}

In the three-band case we have not obtained convergence close to the half-filled situation for $U_d=10.3\;$eV. This is the reason why the plots in Fig. \ref{fig:delt_tab} are drawn only down to $\delta\approx 0.06$. It is not clear if the SC pairing amplitudes drop to zero, as we reach the half-filled situation. In Fig. \ref{fig:delt_Udep} we show the dominant $\Delta_{dd}^1$ amplitude for $\delta=0$ as a function of $U_d$ up to the highest value of $U_d$ for which the convergence could be reached. By carrying out the linear extrapolation in the high-$U_d$ region we estimate that the SC amplitude is completely suppressed above the upper critical value $U_d=U_d^c\approx 13\;$eV.

\begin{figure}
 \centering
 \includegraphics[width=0.5\textwidth]{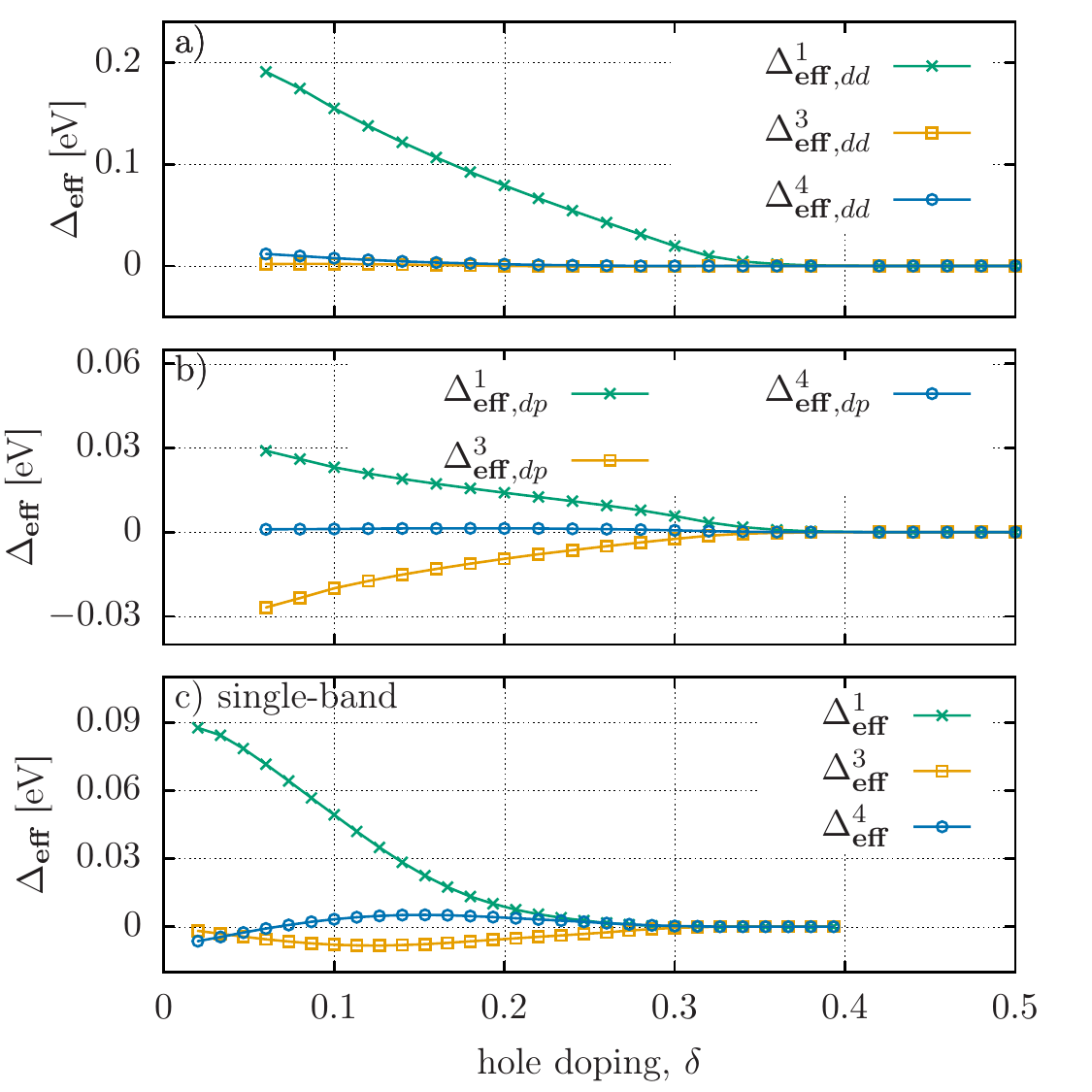}
 \caption{Effective superconducting gaps between $d$-$d$ (a) and $d$-$p$ (b) orbitals as a function of doping for $U_d=10.3\;$eV, $U_p=4.1\;$eV (cf. Fig. \ref{fig:Cu_O_pairing} and Eq. (\ref{eq:SC_amplitudes})). The calculated $p$-$p$ effective gaps are zero. In (c) we show the first, third, and fourth nearest-neighbor effective gap for the case of single band Hubbard model with $t=-0.35\;$eV, $t'=0.25|t|$, and $U=6\;$eV.}
 \label{fig:delt_eff}
\end{figure}

\begin{figure}
\centering
\includegraphics[width=0.5\textwidth]{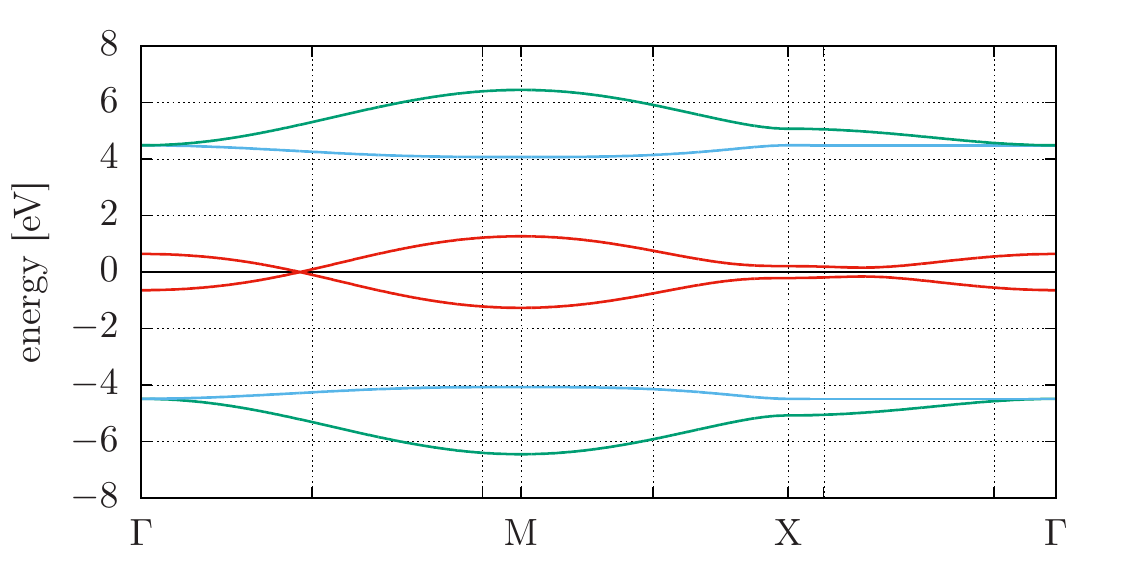}
\caption{Quasiparticle dispersion relations in the SC state for $\delta=0.24$ and for the same model parameters as in Figs. \ref{fig:delt_tab} and \ref{fig:delt_eff}.}
\label{fig:disrel_SC}
\end{figure}

In Fig. \ref{fig:delt_eff} we display the effective SC gaps as a function of doping, both for the three-band (a,b) and single-band cases (c) for the same values of model parameters as those selected in Fig. \ref{fig:delt_tab}. As one could expect, also here the dominant contribution comes from the nearest-neighbor $d$-$d$ pairing. However, in contrast to the pairing amplitudes shown in Fig. \ref{fig:delt_tab}, the dominant effective gap increases as one approaches the half-filled situation. Such a behavior has also been reported for the case of single band Hubbard model analyzed within the VMC approach\cite{Eichenberger_VMC_2007} and for the DE-GWF calculations for the single-band $t$-$J$\cite{Kaczmarczyk2014} and t-J-U\cite{Zegrodnik_3} models. It should be noted that the effective SC gaps within the $p$-orbital sector are zero even though the corresponding pairing correlations have non-zero values (cf. Fig. \ref{fig:delt_tab} c and d). This is because electrons residing on $p$ orbitals are not significantly correlated due to small value of $U_p$ in comparison to $U_d$. Therefore, the pairing correlations between the $p$ orbitals are induced by the appearance of both $d$-$d$ and $d$-$p$ parings, in a manner analogous to the proximity effect in superconductor-normal metal heterostructures. However, such an induced $p$-$p$ pairing correlations do not contribute to the spectral gap meaning that $\Delta_{\mathrm{eff},pp}\equiv 0$. The quasiparticle dispersion relations which result from the effective Hamiltonian (\ref{eq:H_effective}) for the SC phase and for selected value of doping ($\delta=0.24$) are shown in Fig. \ref{fig:disrel_SC}. As one can see the antibonding hybridized band (red solid line) is gapped apart from the nodal point between $\Gamma$ and $M$ due to the $d$-$wave$ symmetry of the SC gap. A similar band structure appears for other dopings. This quasiparticle structure can be compared with that for bare bands depicted in Fig. \ref{fig:disrel_H0}.

\subsection{Overall behavior and phase diagram}
\begin{figure}
 \centering
 \includegraphics[width=0.5\textwidth]{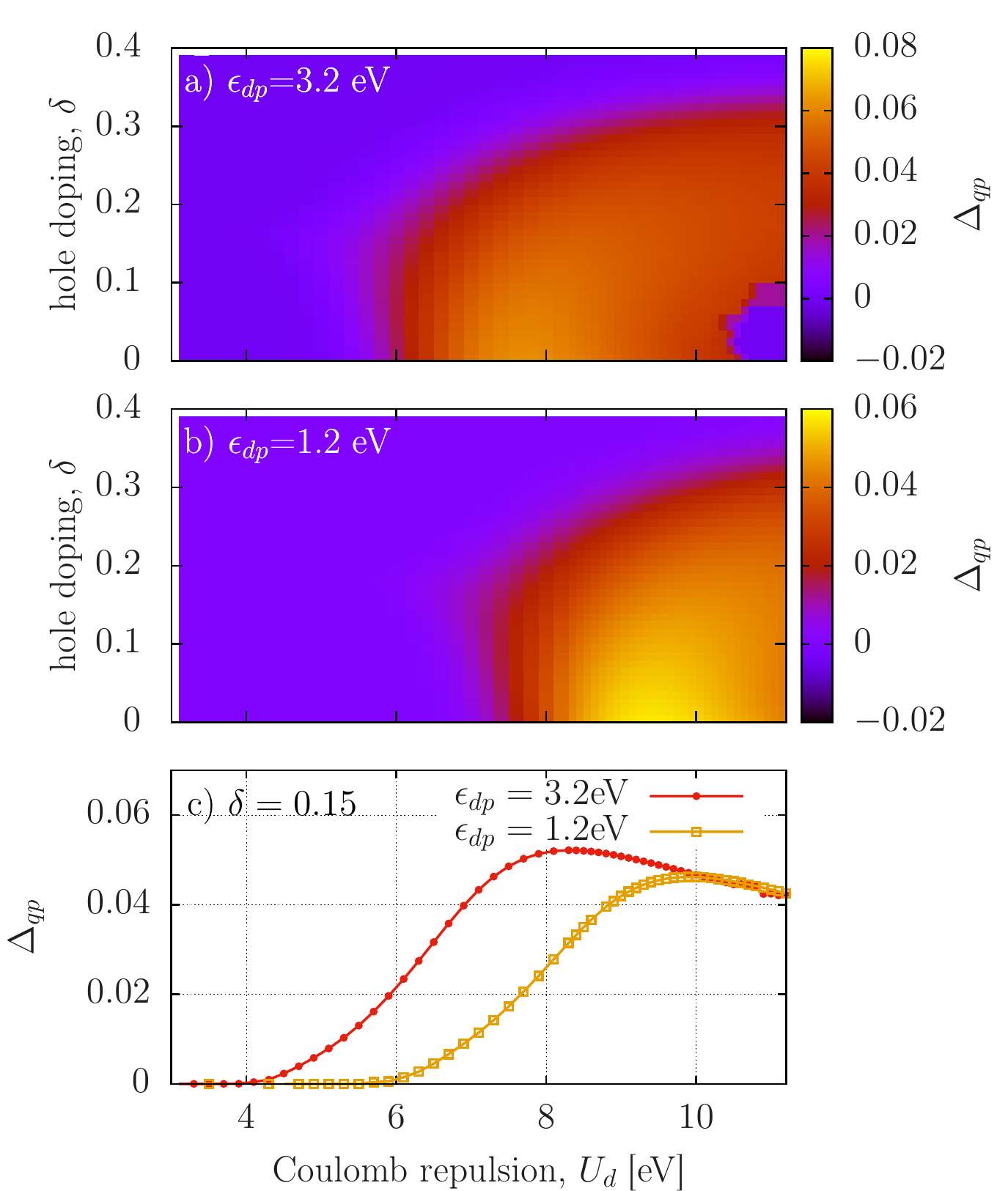}
 \caption{The quasiparticle superconducting gap $\Delta_{qp}$ as a function of doping and Coulomb repulsion on the $d$ orbitals for $\epsilon_{dp}=3.2\;$eV (a) and $\epsilon_{dp}=1.2\;$eV (b); and $U_p=4.1\;$eV. (c) $\Delta_{qp}(U_d)$ plots for $\delta=0.15$ and for the two values of $\epsilon_{dp}$, as for (a) and (b). The irregular region with $\Delta_{qp}=0$ for high values of $U_d$ close to $\delta=0$ in (a), is where we could not achieve convergence of our calculation scheme.}
 \label{fig:delt_diag}
\end{figure}

Next, we turn to the analysis of the question how $U_d$ and $\epsilon_{dp}$ parameters influence the superconducting state. In Fig. \ref{fig:delt_diag} we show the maps of quasiparticle pairing amplitude in $(U_d,\delta)$ space for two selected values of $\epsilon_{dp}$ which differ by 2$\;$eV. Both of them can be regarded as realistic for selected cuprates. Again, the maps resemble those for the single-band Hubbard or t-J-U models (cf. Figs 3 and 4 in Ref. \onlinecite{Zegrodnik_1}), with the paired phase appearing for high enough values of the Coulomb interaction and confined to the region with $\delta\lesssim 0.35$. 

Also, in both single- and three-band cases one can single out three distinct regions: (i) weak-correlation regime (low $U$ or $U_d$) for which the pairing amplitude increases with increasing $U$ or $U_d$; (ii) the intermediate-correlation regime placed around the maximum of the pairing amplitude as a function of $U$ or $U_d$; (iii) strong-correlation regime, with large $U$ or $U_d$ in which the pairing amplitude is decreasing back with the increasing $U$ or $U_d$. For the case of the single band models the intermediate-correlation regime appears close to $U\approx W$, where $W$ is the bare band-width. It is not clear what determines the analogous critical value of $U_d$ in the three-band case. From our analysis we can see that in the three-band case the sequence of the three regimes may be shifted on the $U_d$ axis by changing $\epsilon_{dp}$, which does not have its analog in the single-band case. In Fig. \ref{fig:delt_diag}c we illustrate that effect by drawing $\Delta_{qp}$ vs. $\delta$ for two values of $\epsilon_{dp}$, which differ by $2\;$eV. As one can see, the maximum of $\Delta_{qp}$ as a function of $U_d$ is shifted also by about $2\;$eV. However, the corresponding change of the hybridized antibonding band-width is only $\approx1\;$eV. Such a situation can be understood by looking at the energy change corresponding to the electron transfer from the oxygen atomic site to the nearest-neighbor copper atomic site for the parent compound. It is equal to $\Delta E=U_d-U_p+\epsilon_{dp}$ and corresponds to the lowest energy excitation. The value of $\Delta E$ should be considered as that determining the strength of the electron correlations. By reducing $\epsilon_{dp}$ by $2\;$eV one also reduces $\Delta E$ therefore the strong correlation regime moves by $2\;$ eV towards higher $U_d$ values what is actually seen in Fig. \ref{fig:delt_diag}.

Since the high-temperature superconductors are placed in the strong-correlation regime but close to the intermediate one, the decrease of both $U_d$ and $\epsilon_{dp}$ results in shift towards the intermediate regime, where the higher values of the pairing amplitudes appear. Such a conclusion is in agreement with our analysis of the hole content distribution [$\tilde{n}_d(2\tilde{n}_p)$] according to which for lower values of $U_d$ and $\epsilon_{dp}$ the $\rho$ parameter increases together with the decrease of copper hole-content in favor of the oxygen-hole content for the parent compound. It has been reported experimentally that such changes correspond to enhancement of maximal $T_C$\cite{Rybicki2016}. 



\section{Conclusions}

By analyzing the normal state characteristics we have shown that the correlation effects taken into account by either Gutzwiller- or Jastrow-type variational wave function lead to kinks in the orbitally resolved double occupancy and electron concentration for the half-filling in the three-band $d$-$p$ model (cf. Fig. \ref{fig:PM_characteristics}), which have also been reported by other methods\cite{DMFT_3band_2008,QMC_3band_2016} dedicated to strongly correlated systems, and do not appear in the Hartree-Fock approximation. In this respect, the correlations alter significantly the role of $\rho$ parameter, reducing it in the hole doped regime and increasing its value for the case of electron doping (cf. Fig. \ref{fig:PM_npnd}). Therefore, the low values of $\rho$ measured in the hole doped cuprates with the lowest experimentally determined $\rho\approx 0.2$ (for La-214)\cite{Rybicki2016} should be considered as a signature of strong electron correlations. The values of $\rho$ obtained here ($\rho\approx0.7$-$1.0$) within DE-GWF/VMC calculations correspond to the Bi-, Hg-, Tl-based cuprate compounds\cite{Rybicki2016}.

We have analyzed the paired phase in the three-band $d$-$p$ model with the use of the DE-GWF method and have shown, that due to the electron correlation effects taken into account within the variational approach, the SC state is stable in the doping region $\delta\lesssim 0.35$, with the maximal value of the dominant $d$-$d$ pairing amplitude appearing for $\delta\approx0.19$ (optimal doping). Those values correspond well with those determined in the single band calculations\cite{Zegrodnik_1,Kaczmarczyk2014} and in numerous experimental situations for the cuprates\cite{cuprates_rev_2006}. 

Within the three-band case the dominant contribution to the SC state comes from the pairing between the nearest-neighbor $d$-$d$ atomic sites, which also reflects promarily the behavior of the quasiparticle gap for the antibonding hybridized band. The calculated pairing amplitudes between the copper atomic sites resemble those corresponding to the subsequent nearest neighbors of the square lattice for the SC phase in the single-band Hubbard model (cf. Figs. \ref{fig:delt_tab} a and b). Such a connection is also seen in the calculated effective gaps (cf. Figs. \ref{fig:delt_eff} a and c). Another similarity between the single and three-band models is the characteristic behavior of the pairing amplitude as a function of both $\delta$ and $U_d$ ($U$ for the single-band case) with the weakly-, intermediate-, and strongly-correlated regimes visible (cf. Fig. \ref{fig:delt_diag} here and Figs. 3 and 4 in Ref. \onlinecite{Zegrodnik_1}). It should be noted, that the close relation between the relevant subbands of the three- and single-band models has been reported in Ref. \onlinecite{Avella2013} with the use of composite operator method. However, that analysis did not include the paired states. On the other hand, the differences between the single- and three-band picture of the cuprates with respect to the strength of spin-fluctuations and their relation to the pairing mechanism, have been singled out in Refs. \onlinecite{Ebrahimnejad2014,Ebrahimnejad2016}.

The charge transfer energy ($\epsilon_{dp}$), which does not have its correspondent in the single-band case, tunes the strength of correlations. Namely, for smaller $\epsilon_{dp}$ values the correlations seem to be suppressed what means that stronger Coulomb repulsion ($U_d$) is necessary to induce the SC state. Also, since the copper-based materials are placed in the strongly correlated regime close to the intermediate one, it results from our analysis that by decreasing $\epsilon_{dp}$ and $U_d$ one moves towards the intermediate-correlations regime, where the values of pairing amplitudes are higher. This in turn may lead to a higher critical temperature. Such a conclusion is in agreement with our analysis of the hole content distribution and the experimental findings, according to which the reduction of the hole content at copper in favor of oxygen in both the parent compound and hole-doped situation leads to an enhancement of the maximal critical temperature\cite{Rybicki2016}. At the present stage of our research the agreement with the mentioned experiments is only qualitative, since we were not able to fit directly to the measured copper and oxygen hole contents and obtain the changes of pairing amplitudes which would correspond to the reported T$_C$ for different cuprate compounds (cf. Fig. 2 in Ref. \onlinecite{Rybicki2016}). For example, the experiments for La-214 report $\rho\sim 0.2$, and such low values cannot be reproduced in our theoretical approach within the range of realistic model parameters. The fact that there is correlation between the apical oxygen distance and the value of $\rho$, suggests that to achieve the quantitative agreement between theory and experiment in this respect one should include the orbitals originating from those apical oxygen states in a manner presented in Ref. \onlinecite{4band_apical}.



At the end, it should be noted that the VMC calculations have been carried out for limited system consisting of the 4x4 CuO$_2$ complexes, whereas the DE-GWF method allows for analysis of infinite systems. Also, within the DE-GWF approach, we included only onsite correlation operator acting on the copper atomic sites, whereas within the VMC calculations more involved wave function has been applied with the intersite Jastrow factors. In spite of those differences, the agreement between the two methods is very good (cf. Figs. \ref{fig:PM_characteristics}, \ref{fig:PM_npnd}). This last feature speaks again for the dominant role of the Cu $d$ electrons.

\section{Acknowledgement}
This research has been financed through the Grant SONATA, No. 2016/21/D/ST3/00979 from the National Science Centre (NCN), Poland.

\appendix
\section{}

\begin{figure}
 \centering
 \includegraphics[width=0.5\textwidth]{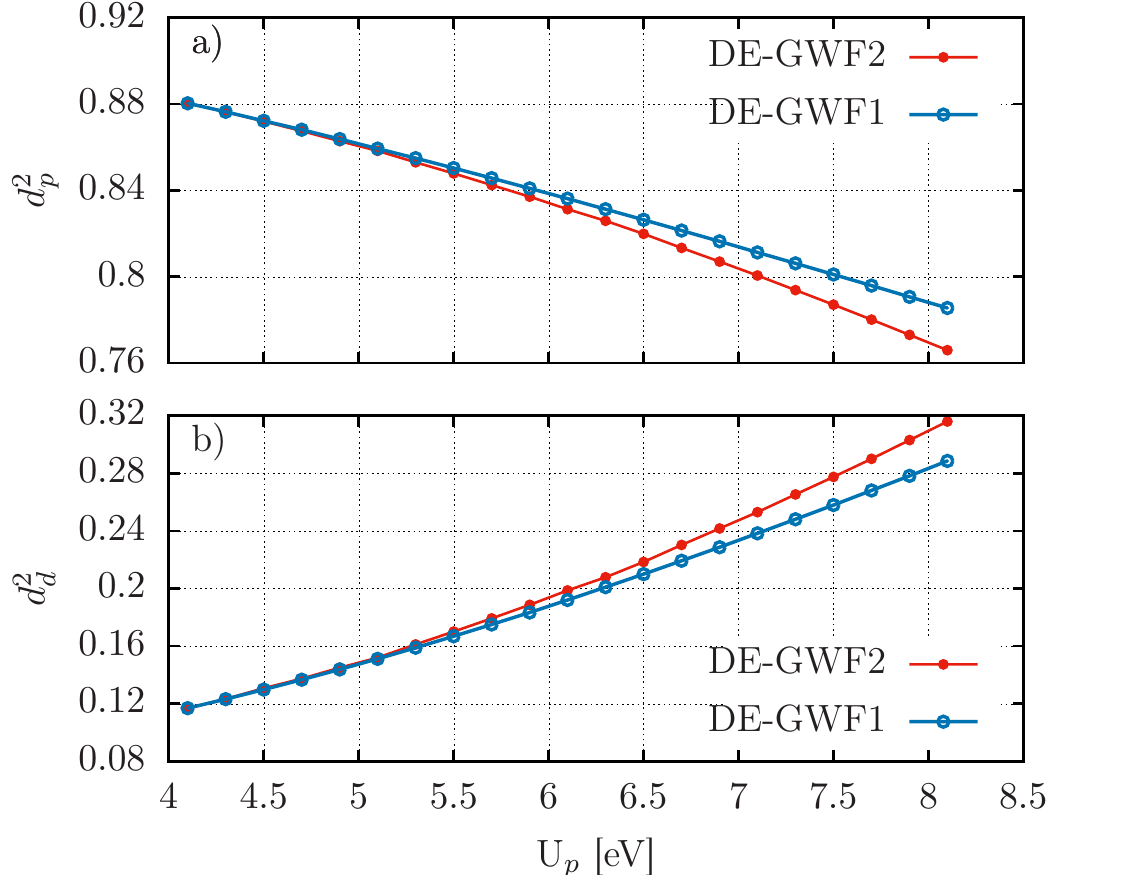}
 \caption{Double occupacies on the oxygen (a) and copper (b) atomic sites as a function of $U_p$ for $U_d=7.85\;$eV and $\delta=0.2$ calculated by two variants of the DE-GWF scheme. In the first one (DE-GWF1) we fix $x_p\equiv0$ and minimize only over $x_d$, whereas in the second (DE-GWF2) the full minimization over both $x_d$ and $x_p$ is carried out.}
 \label{fig:Updep}
\end{figure}

Here we show that in the parameter regime significant for the cuprates it is justified to apply the approximation for which $x_p\equiv 0$. In such a situation the correlation operator from Eq. (\ref{eq:P_Gamma}) acts only on the copper atomic sites, what simplifies significantly the calculations. Nevertheless, the electron correlations; which result from the dominant interaction ($U_d$) in the system are taken into account by minimizing the system energy over $x_d$. In Fig. \ref{fig:Updep} we show the double occupancies on the copper and oxygen atomic sites as a function of $U_p$ calculated within DE-GWF method by assuming either $x_p\equiv 0$ (DE-GWF1), or obtained by using the complete form of the correlation operator (\ref{eq:P_Gamma}), i.e., minimization over both $x_d$ and $x_p$ (DE-GWF2). As one can see, the differences between the two calculation schemes become visible as one increases the value of $U_p$. Nevertheless, for $U_p\approx 4-6\;$eV, appropriate for the cuprates, the results practically coincide. 

%

\end{document}